\documentclass[aip,jmp]{revtex4}
\usepackage{amsmath, amsthm, amssymb}%
\usepackage{color}%
\usepackage{comment}%
\usepackage{url}%
\usepackage{graphicx}%
\usepackage{epsfig}%
\usepackage{docs}%
\usepackage{bm}%
\usepackage[colorlinks=true,linkcolor=blue]{hyperref}%
\expandafter\ifx\csname package@font\endcsname\relax\else
 \expandafter\expandafter
 \expandafter\usepackage
 \expandafter\expandafter
 \expandafter{\csname package@font\endcsname}%
\fi
\begin{document}

\title{A quantum mechanical well and a  derivation of a $\pi^2 $ formula.}%

\author{Anna Lipniacka, Bertrand Martin Dit Latour}%
\email{anna.lipniacka@uib.no,bertrand.martindl@cern.ch}%
\affiliation{Department of Physics and Technology, University of Bergen\\ All{\'e}gaten 55 \\5007 Bergen, Norway}%

\date{October 2017}%
\revised{xxx 2017}%

\begin{abstract}
\noindent
%===================> note abtract     =====> To be filled <=====%

Quantum particle bound in an infinite, one-dimensional square potential
well is one of the
problems in Quantum Mechanics (QM) that most of the textbooks start from.
There, calculating an allowed  energy spectrum for an arbitrary wave
function often
involves Riemann zeta function resulting in a $\pi$ series  \cite{griffits-1}.
In this work, two ``$\pi$ formulas'' are derived when
calculating a spectrum of  possible outcomes of  the   momentum
measurement for  a particle confined in such a 
well, the  series,
$\frac{\pi^2}{8} = \sum_{k=1}^{k=\infty}  \frac{1}{(2k-1)^2}$,
and the  integral $\int_{-\infty}^{\infty} \frac{sin^2 x}{x^2} dx =\pi$.
%%%%%This spectrum appears to be different than 
%naively expected.
The  spectrum  of the momentum operator
appears to peak on classically allowed momentum values
only for the states with even quantum number.
The present article 
is inspired by another quantum mechanical derivation
of $\pi$ formula in \cite{wallys}.
%mention here calculation in Griffith and say this is perhaps not the only
%way..

%=========================================================================%

\end{abstract}

\maketitle

%\section{Introduction \label{sec:intro}}

The $\pi^2$ series  :\\

\begin{equation} \label{eq:form1}
  \frac{\pi^2}{8} = \sum_{k=1}^{k=\infty}  \frac{1}{(2k-1)^2}= 1+\frac{1}{3^2}+\frac{1}{5^2}...
\end{equation}

\noindent
cited for example in \cite{numbers} and
\cite{wolfram} is not attracting much attention,
perhaps due to its relatively slow convergence.
The  integral
$\int_{-\infty}^{\infty} \frac{sin^2 x}{x^2} dx =\pi$ can be calculated
using complex number analysis.
As  shown here,  both of these have to be true to ensure
the consistency of QM formalism when calculating
the spectrum of possible outcomes  of the momentum measurement  
for  a quantum particle in a one-dimensional (1D) infinite square well.
The derivation involves  Fourier
series and Fourier transform.
The momentum operator spectrum appears to be  different than  naively expected.
A derivation of the 
Wallis formula for   $\pi$  in the context of
QM analysis of hydrogen atom  was demonstrated in \cite{wallys}.\\

%\noindent
%In the following, the formula  \ref{eq:form1} and the result
%$\int_{-\infty}^{\infty} \frac{sin^2 x}{x^2} dx =\pi$ 
%and \ref{eq:form3}
%will be derived \\
\vspace{0.3cm}

%\noindent
%\section{Momentum operator  eigenvalues in an infinite 1D square well \label{sec:derivation}}

%%%%%%%%%%%%%%%%%%%%%%%%%%%%%%%%%%%%%%%
\begin{comment}
\noindent
The  one-dimensional square well is placed between $0<x<L$.\\

\noindent
$E \Psi(x)  = [  - \frac{{\hbar  }^2}{ 2m }  \frac{d^2}{dx^2}  + V(x)  ] \Psi(x)\\
- \frac{2m}{\hbar^2} (E -V(x)) \Psi(x)  =  \frac{d^2}{dx^2} \Psi(x)\\
\frac{2m}{\hbar^2} (V-E) = \frac{ \Psi''}{ \Psi}\\
$\\

\noindent
We have the well with potential infinite beyond 0 and x=L thus:\\
$V (x) = 0 ~ for~ 0 <x<L $  and  $V(x)= \infty$  otherwise. 
The wave function inside the well is the solution of free SE.\\

\vspace*{1cm}

\noindent
$- \frac{\hbar^2} {2m} \frac {d^2} {d x^2} \Psi (x)  = E \Psi (x) $\\

\noindent
 $\Psi (x)  = A sin (x \frac{\sqrt{2mE}} {\hbar} ) +B cos (x \frac{\sqrt{2mE}}{\hbar} )$ \\
 $\Psi (0)  =0 ~ \rightarrow B=0$ \\ 
 $\Psi (L)  = A sin (L \frac{\sqrt{2mE}} {\hbar} )= 0 ~ \rightarrow  ~ L \frac{\sqrt{2mE_n}}{\hbar}= n\pi$ \\

\noindent
$ E_n =   \frac{1}{2m}  ( \frac{ n\pi \hbar} {L})^2 = \frac{<p^2>}{2m} $ \\

\noindent
 $\Psi_n (x)  = A sin ( \frac{x} {L} n\pi )$  and  
$\int_{0}^{L} { |\Psi(x)|}^2 = 1 ~ \rightarrow  ~ 
A = \sqrt {\frac{2}{L}}$\\

\end{comment}

%%%%%%%%%%%%%%%%%%%%%%%%%%%%%%%%%%%%%%%%%%%%%%%

\noindent 
It is a fundamental assumption of QM, that most of textbooks
start from  that all information about a system at a given instant of time
can be derived from the wave function \cite{mandl1.1}, $\Psi(\vec r,t)$.
Hermitian operators representing the
measurable quantities (observables) provide the way of deriving
this information. While the average result of a measurement
obtained on an ensemble of identically prepared systems can be calculated
as an expectation
value of an operator  in question, $ \hat{O} $, 
 \begin{equation}\label{formxaa}
  <\hat{O}> = \int \Psi^{*}(\vec r,t) \hat{O} \Psi(\vec r,t) d^3 r
\end{equation}
 the outcomes of  single measurements are
 eigenvalues of this operator.
 The link between   the eigenvalues and the expectation values is provided
 by the formulas below.  For a discrete spectrum of eigenvalues, $o_k$, one gets,

 \begin{equation}\label{formxab}
  <\hat{O}> = \Sigma_k|c_k|^2 o_k
\end{equation}

\noindent 
 whereas the sum is replaced by an integral in case of a continuous spectrum
 of eigenvalues, $k$,

 \begin{equation}\label{formxbb}
  <\hat{O}> = \int |c(k)|^2 k dk
 \end{equation}

\noindent 
The modulus squared
of a coefficient $|c_k|^2$ (the function  $|c(k)|^2dk$ )
represent the  probability  to measure a given eigenvalue $o_k$
(an eigenvalue between $k$ and $k+dk$) in case of a discrete
(continuous) spectrum of eigenvalues.
These coefficients can be calculated when representing
the wave function of the system as a
linear combination of the orthonormal  eigenfunctions of the operator in question. 

\noindent 
%For the states with  defined energy (stationary states) the wave function of the
%system is an eigenfunction of the Hamiltonian operator.
One  typical QM exercise
illustrating the above mechanism is to represent  an  eigenfunction
the  Hamiltonian , $ \Psi(\vec r)$,
as a linear combination of orthonormal  eigenfunctions, $\Phi_k(\vec r)$, of another observable of interest,
$ \Psi(\vec r)= \Sigma_k c_k \Phi_k (\vec r)$
(or $ \Psi(\vec r)= \int c(k) \Phi_k (\vec r)dk$ for continuous spectrum)
in order to calculate
what are the possible outcomes
of the measurements of this  observable
for a state with a well defined energy (a stationary state).
An expansion coefficient $c_{\xi}$ or a function $c(\xi)$
can be calculated as a  scalar product
of $ \Phi^{*}_{\xi}$ and $\Psi(\vec r)$:

\begin{equation}
\label{formabb} c(\xi) = \int  \Phi^{*}_{\xi} ( \vec r ) \Psi  (\vec r) d^3r~~ \text{or} ~~ c_{\xi} = \int  \Phi^{*}_{\xi} ( \vec r ) \Psi  (\vec r) d^3r \\
\end{equation}

\noindent
This exercise, performed below  for the
imaginably simplest QM system with an endeavor to calculate
the possible outcomes
of momentum measurement, yields a somewhat unexpected conclusion.\\

\noindent
The eigenfunctions of the Hamiltonian,

\begin{equation}\label{form6a}
  \hat{H} = - \frac{\hbar^2}{2m} \frac{d^2}{dx^2} + V(x)
\end{equation}

\noindent
inside an infinite one-dimensional  square well 
situated in   $ 0 \le x \le L$  can be found in any QM textbook \cite{griffith-2} and are of the form:\\

\begin{equation}\label{form6}
\Psi_n (x)  = \sqrt {\frac{2}{L}} sin ( \frac{x} {L} n\pi )  
\end{equation}

\noindent
These functions are  orthonormal  inside the well $0 \le x \le L$ and form
a complete set.
They  fulfill correct boundary conditions, disappearing at the borders
of the well since the wave function has to be continuous and cannot exist
in the area of the infinite potential.
They also  give the  known energy spectrum,

\begin{equation}\label{form6ener}
 E_n =   \frac{1}{2m}  ( \frac{ n\pi \hbar} {L})^2 = \frac{<p_x^2>}{2m} 
\end{equation}

\noindent
However, they are not eigenfunctions
of the 1D  momentum operator,  $\hat{p_x} \equiv \frac{\hbar}{i} \frac{ d }{ d x}$.
%Since $\hat {p_x}$ commutes
%with the Hamiltonian inside the well
%the common set of eigenfunction for these two operators  must exists, even if
%the state of a confined particle cannot be the momentum operator eigenfunction.\\
A classical particle bound in the   well
with the kinetic energy $E_n$ would be bouncing back and forth with
the momentum $p_x=\pm \frac{ n\pi \hbar} {L}$. A quantum particle has
$<\hat p^{2l+1}_x>=0$ and  $ <\hat p^{2l}_x> = (\frac{ n\pi \hbar} {L})^{2l}$ as  can
be easily calculated using  space representation of the $\hat{p_x}$ and the
Hamiltonian eigenfunctions with use of the formula \ref{formxaa}.\\
%$\Delta p_x = \frac{ n\pi \hbar} {L}$ \\

\noindent
In order to calculate the  possible outcomes of momentum measurements
one has to use an orthonormal and complete set of the  momentum
operator eigenfunctions.
The usual procedure
is to propose the eigenfunctions with either a  continuous spectrum
of eigenvalues, as suggested in \cite{griffithp22}, normalized
to the Dirac $\delta$ or with  a  discrete spectrum of eigenvalues, normalized inside
the well and fulfilling the periodic boundary conditions \cite{mandl-momentum}.\\

\noindent
For example,
$ \Phi_k(x) = \frac{1}{\sqrt{2 \pi}} exp({ikx})$ is a continuous spectrum
eigenfunction of  ${\hat{p_x}/\hbar}$ normalized as follows,
$ \int_{-\infty}^{\infty} \Phi_k(x) \Phi_k'(x)dx= \delta({k-k'}) $.\\

\noindent
For the discrete spectrum, the  orthonormal
set of momentum eigenfunctions  normalized inside the well,
$ \Xi(x) = \frac{1}{\sqrt{L}} exp({\frac{ipx}{\hbar}})$,  
needs to  fulfill the  periodic boundary
conditions, $ \Xi(L)= \Xi(0)$. These  conditions:\\

\begin{equation}\label{form8}
\frac{1}{\sqrt{L}} exp({\frac{ipL}{\hbar}})=\frac{1}{\sqrt{L}}\\  
\end{equation}

\noindent
result in:\\

\begin{equation}
\label{form9} exp({\frac{ipL}{\hbar}})= 1,  ~ \frac{pL}{\hbar}=2l\pi,\\
\end{equation}

\noindent
where $l$ is an integer number.  \\  

\noindent
Denoting the momentum
eigenfunctions with positive and negative $l$ as follows:\\
 
\begin{equation}
 \label{form10} \Xi^+_n(x) = \frac{1}{\sqrt{L}} exp({\frac{+i2n\pi x}{L}})~,~~~~\Xi^-_n (x) = \frac{1}{\sqrt{L}} exp({\frac{-i2n\pi x}{L}})\\
\end{equation}

%\begin{equation}
%\label{form11}  
%\end{equation}

\noindent
one has now $n$ a natural number, or zero. For $n=0$
one gets  , $\Xi_0 (x) = \frac{1}{\sqrt{L}}$ with zero eigenvalue.  \\

\noindent
%Only ``every second'' of the functions in listed in Appendix A2 (\ref{form7}, \ref{form7a}) enters the
%orthonormal set and
The momentum operator eigenvalues are thus of the form:
${\frac{2n\pi \hbar}{L}}$, different than naively expected, eigenvalues
which are odd number multiplications of ${\frac{\pi \hbar}{L}}$ being absent.
Since these eigenvalues can in principle
be measured the question of what is the momentum operator  spectrum is not
purely academic one. 
For completeness, the orthonormality of the functions $\Xi^{\pm}_n(x)$
is shown in  the Appendix A2.\\

\noindent
 Probabilities to measure  given $p_x$ values 
 are calculated below and compared for continuous and discrete
 spectra. 
The probability, $P_n(k)dk=|c_n(k)|^2dk$
to measure a given value of
$k$, $p_x=\hbar k$,  for a given state
$\Psi_n(x)$ of the particle in the well,
using the continuous spectrum momentum eigenfunctions can
be calculated as follows,\\

\begin{equation}
\label{formbb} c_n(k) = \int_0^L  \Phi^{*}_k ( x' ) \Psi_n  (x') d x'= \frac{1}{\sqrt{2 \pi}} \sqrt {\frac{2}{L}} \int_0^L  exp({-ikx}) sin ( \frac{n\pi x} {L}  )  d x  \\
\end{equation}

\begin{equation}
\label{formaa} P_n(k)= |c_n(k)|^2 = \frac{n^2\pi L} {(L\cdot k +n \pi)^2(L \cdot k-n\pi)^2}|(-1)^n exp(iLk)-1|^2  \\
\end{equation}

\noindent
Since $|(-1)^n exp(iLk)-1|^2= 4 cos^2(Lk/2) $ for odd  $n$ and
$|(-1)^n exp(iLk)-1|^2= 4 sin^2(Lk/2) $ for even $n$ one gets 
a somewhat surprising result that the probability
to measure a certain momentum $p_x=\hbar k$  peaks at the classical
momentum values $p_x= \hbar k = \pm \frac{n \hbar  \pi }{L}$ only for even $n$ values,
whereas for odd $n$ values it peaks up at $p_x=0$. Figure 1 shows the
probability density as
a function of a  dimensionless variable, $\xi= kL= \frac{p_x L}{\hbar}$
proportional to the
momentum eigenvalue, for the ground state ($n=1$) and the first excited state ($n=2$) of the well.
The expectation value of
the momentum operator is zero, since the probability density
in formula \ref{formaa} is a  symmetric function.
For this momentum space wave function to be correctly normalized and to
give the correct expectation value of$ <\hat p^{2}_x> = (\frac{ n\pi \hbar} {L})^{2}$ the
following 
must hold (for $n=1$):\\

\begin{equation}
\label{formba} \int_{-\infty}^{\infty}|c_1(\xi=kL)|^2 d \xi = \int_{-\infty}^{\infty} \frac{4 \pi} {(\xi + \pi)^2(\xi-\pi)^2} cos^2(\xi/2) d\xi = 1  \\
\end{equation}

\begin{equation}
\label{formaab} \int_{-\infty}^{\infty}|c_1(\xi=kL)|^2\xi^2 d \xi = \int_{-\infty}^{\infty} \frac{4 \pi} {(\xi + \pi)^2(\xi-\pi)^2} cos^2(\xi/2)\xi^2 d\xi = \pi^2  \\
\end{equation}

\noindent
These integrals are equivalent  to   $\int_{-\infty}^{\infty} \frac{sin^2 x}{x^2} dx =\pi$, see the  Appendix A1 for the algebra.
Note that  it is not
possible to sensibly  calculate the expectation value of the higher  powers,
$<\hat p^{2l}_x>$, of
the momentum operator using the momentum space wave function $c_n(k)$,
the integral is divergent already for $l=2$.\\

\begin{figure}[htb]
\begin{center}
\begin{minipage}[t]{0.48\linewidth}
\mbox{\epsfig{file=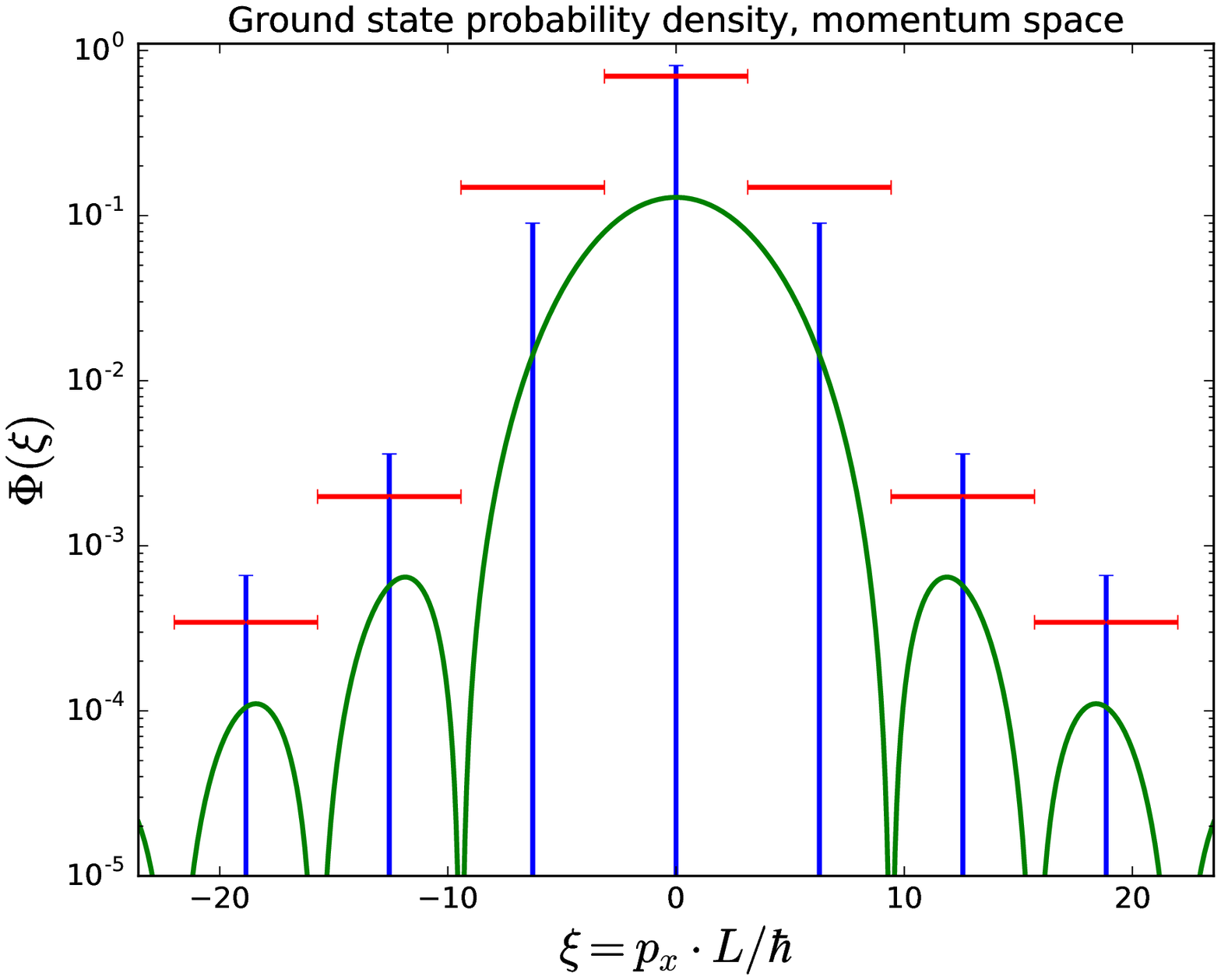,width=\linewidth}}
\end{minipage}
\hfill
\begin{minipage}[t]{0.48\linewidth}
\mbox{\epsfig{file=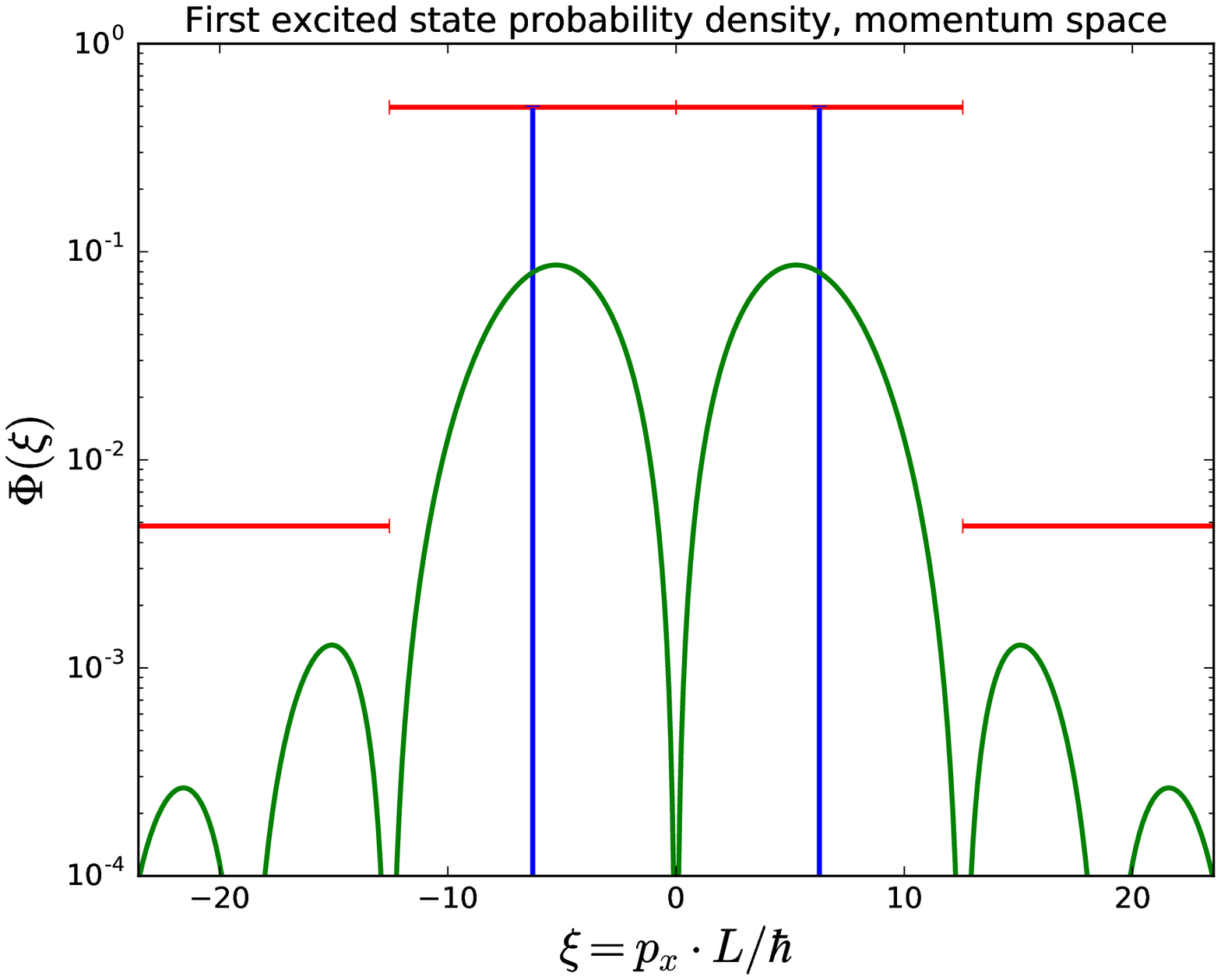,width=\linewidth}}
\end{minipage}
\end{center}
\caption{Probability density (green line)  and discrete probabilities (blue columns)
  to measure a given momentum value for
  the ground state, left, and the first excited state, right, for a particle in a one dimensional
  infinite well of the size $L$.  Red bars show  values of integrals of the probability density in the
  range of the bar.}
\label{fig:g0}
\vfill
\end{figure}

\noindent
The possible spectrum of outcomes of momentum measurements for a
particle in the ground state  and the first excited state is  obtained below using the 
momentum operator eigenfunctions with the discrete spectrum.
Here one needs to express 
the  eigenfunctions of the Hamiltonian \cite{mandl2},
$\Psi_1 (x)  =  \sqrt {\frac{2}{L}}  sin ( \frac{\pi x} {L}  )$
and  $\Psi_2 (x)  =\sqrt {\frac{2}{L}} sin ( \frac{2\pi x} {L}  )$,  
as a combination of momentum operator
eigenfunctions in  formula \ref{form10}.\\

\noindent
Finding the expansion coefficients
for the first excited state eigenfunction in the well is
straightforward because 
$\Psi_2(x)$  is a
simple combination of
$\Xi^+_1 = \frac{1}{\sqrt{L}} exp({\frac{+i2\pi x}{L}})$   and $\Xi^-_1 = \frac{1}{\sqrt{L}} exp({\frac{-i2\pi x}{L}})$,
$\Psi_2(x)=\frac{i}{\sqrt{2}}(\Xi^-_1 -\Xi^+_1 )$.
Thus there are two possible results of the  momentum measurement, $p_x= \frac{2 \pi \hbar}{L}$ and $p_x= -\frac{2 \pi \hbar}{L}$,  occurring
with equal probabilities, resulting in the $< \hat{p_x}>=0$, in agreement with
the  classical
result.\\

\noindent
The result for the ground state wave function
$\Psi_1 (x)  =  \sqrt {\frac{2}{L}}  sin ( \frac{\pi x} {L}  )$ is 
more involved.
The  expansion coefficients can be calculated using the formula \ref{formabb}:\\  

\begin{equation}
\label{form12} \Psi(x ) =   \Sigma_i { c_i   \Xi_i  (x)  } ~, ~~ c_n = \int_0^L  \Xi^*_n ( x' ) \Psi  (x') d x' \\
\end{equation}

\noindent
For the  ground state wave function
$\Psi_1(x)  = \sqrt {\frac{2}{L}} sin ( \frac{x\pi} {L}  )$,     
the expansion coefficients are:\\

\begin{equation}
\label{form15}  c_{\pm k} = \int_0^{L}  \Xi^*_{\pm k} ( x' ) \sqrt {\frac{2}{L}} sin ( \frac{x'\pi} {L}  ) d x' \\
\end{equation}
  
\begin{equation}
\label{form18} c_{\pm k} =  \frac{\sqrt 2}{2i} \left [  \frac{exp{(\frac{ i (\mp 2k+1)\pi x'}{L}})}{i\pi(\mp 2k+1)} - \frac{exp{( \frac{i (\mp 2k -1)x'\pi} {L} })}{i\pi(\mp 2k-1)} \right ]_0^L
\end{equation}

\noindent
Since both, $\mp 2k+1$, and, $\mp 2k-1$, are odd numbers 
both exponential functions are $=-1$ for $x'=L$. One gets:\\

\begin{equation}
\label{form21}  c_{+k} =  - \frac{1}{\pi} \cdot \frac{2\sqrt {2}}{(2k-1)(2k+1)} =c_{-k}  
\end{equation}

\noindent
The modulus squared of a given coefficient defines the
probability to measure a given momentum value in the ground
state of the infinite square well:\\

\begin{equation}
 \label{form22}  |c_{\pm k}|^2 =  \frac{1}{\pi^2} \cdot  \frac{8}{(2k-1)^2(2k+1)^2 }  
\end{equation}
  
\noindent
All the coefficients are  non-zero, thus
the whole  spectrum of the  momentum operator eigenvalues can be
measured for a particle in the ground state of the
infinite square well. Figure 1 shows the numerical values of some of these coefficients
compared with the probability density in formula \ref{formba} calculated using the
continuous spectrum of momentum eigenvalues, for
the ground state and the first excited state of the well.
In the ground state, the largest probability is to measure  null momentum,
the same conclusion was reached using the momentum eigenfunction with
the continuous spectrum of eigenvalues.
Again, as  expected, $< \hat{p_x}>=0$. \\

%%%%%It is interesting to  calculate explicitly some of them:\\

\noindent
%$|c_0|^2=\frac{8}{\pi^2} \simeq 0.81$ , $|c_{\pm 1}|^2 = \frac{8}{9\pi^2} \simeq 0.09 $,
%$|c_{\pm 2}|^2 = \frac{8}{225\pi^2} \simeq 0.004$\\

\noindent
The   moduli squared of all coefficients  in formula  \ref{form22}  must sum up to unity,  since
they together represent the probability of measuring any momentum value. Thus:

\begin{equation}
  |c_0|^2 + 2 \cdot \sum_1^{\infty}|c_k|^2=  \frac{8}{\pi^2}+  \frac{2}{\pi^2}   \cdot \sum_1^{\infty}  \frac{8}{(2k-1)^2(2k+1)^2 }  =1
\end{equation}

\noindent
This implies the following
relation  involving $\pi^2$ to be true, identical to the equation \ref{eq:form3} in Appendix A1: \\

\begin{equation}
\label{form23} \frac{1}{2}+ \sum_1^{\infty} \frac{1}{(2k-1)^2 (2k+1)^2 }   = \frac{\pi^2}{16} 
\end{equation}

\noindent
The eigenvalues  of the  Hamiltonian, $\hat{H}=\frac{\hat p^2}{2m}$   for  the particle in the well are given in the formula \ref{form6ener}.
%$ E_n =   \frac{1}{2m}  ( \frac{ n\pi \hbar} {L})^2 = n^2 E_1 $.
The expectation
value of the Hamiltonian on the ground state eigenfunction,
$\Psi_1(x)  = \sqrt {\frac{2}{L}} sin ( \frac{x\pi} {L}  ) $,
is  equal to  $E_1$. If one  calculates the expectation
value of $\hat{H}$ in \ref{form6a}  using the  expansion in formula
\ref{form12} one gets:\\

\begin{equation}
E_1= < \hat{H}>  =  2* \sum_1^{\infty}  (2k)^2 E_1  \cdot  \frac{1}{\pi^2}  \cdot  \frac{8}{(2k-1)^2 (2k+1)^2 } \\
\end{equation}
    
\noindent
Thus another  relation involving $\pi^2$ in \ref{eq:form2} in Appendix A1 has to be fulfilled :\\

\begin{equation}
\label{form24} \sum_1^{\infty}  \frac{(2k)^2 }{(2k-1)^2 (2k+1)^2 } = \frac{\pi^2}{16}  
\end{equation}

\noindent
The formulas  in \ref{form23} and  in \ref{form24} are trivially
equivalent to each other and to  the  known $\pi^2$ series in the
formula \ref{eq:form1}, see Appendix A1.
It has been thus demonstrated that
$\pi^2$ formula in \ref{eq:form1} stems from the spectrum of possible
outcomes of momentum measurements for a QM particle confined in a one-dimensional box.
The continuous and discrete spectra of momentum eigenvalues show similar features
peaking at $p_x=0$ for the ground state (and any odd $n$) and at the classically allowed
momentum values for the first excited state (any even $n$).
The second power of momentum operator is the highest even power the expectation value of which can be
sensibly calculated  with the 
momentum representation
of the infinite well wave function, both  in its continuous $c(k)$ or
discrete $c_k$ form.\\

\noindent
\section{Acknowledments}
\noindent

We thank Per Osland for cross-checking the formulas and pertinent comments on the
presentations of these results.\\

\noindent
\section{Appendix A1}
\noindent

\noindent
The equations \ref{formba} and \ref{formaab} can be rewritten as follows:\\

\begin{equation}
\label{formbaz} \int_{-\infty}^{\infty} \frac{4 } {(z + 1)^2(z-1)^2} cos^2(z\pi/2)  dz = \pi^2  \\
\end{equation}

\begin{equation}
\label{formaabz} \int_{-\infty}^{\infty} \frac{4 } {(z + 1)^2(z-1)^2} cos^2(z\pi/2)z^2 dz = \pi^2  \\
\end{equation}

\noindent
Adding them and dividing by two one obtains:\\

\begin{equation}
\label{formaabz1} \int_{-\infty}^{\infty} (\frac{1 } {(z + 1)^2} + \frac{1}{(z-1)^2}) cos^2(z\pi/2) dz = \pi^2  \\
\end{equation}

\noindent
or, after  elementary integration variable changes:\\

\begin{equation}
\label{formaabz2} \int_{-\infty}^{\infty} \frac{cos^2(y\pi/2-\pi/2) } {y^2}dy + \int_{-\infty}^{\infty} \frac{cos^2(y\pi/2+\pi/2)}{y^2}  dy = \pi^2  \\
\end{equation}

\begin{equation}
\label{formaabz3} 2 \int_{-\infty}^{\infty} \frac{sin^2(y\pi/2) } {y^2}dy = \pi^2  \\
\end{equation}

\begin{equation}
\label{formaabz4}  \int_{-\infty}^{\infty} \frac{sin^2(x) } {x^2}dx = \pi  \\
\end{equation}

\noindent
A simple  rearrangement of  the equation \ref{eq:form1} leads to the  two formulas below:\\

\noindent
\begin{equation}\label{eq:form2}
SUM1 \equiv  \sum_1^{\infty}  \frac{(2k)^2 }{(2k-1)^2 (2k+1)^2 } = \frac{\pi^2}{16}
\end{equation}

\noindent
and, \\

\begin{equation}\label{eq:form3}
\frac{1}{2} + SUM2 \equiv \frac{1}{2}+ \sum_1^{\infty} \frac{1}{(2k-1)^2 (2k+1)^2}   = \frac{\pi^2}{16}   \\
\end{equation}
  
\noindent
The  equations \ref{eq:form2}, \ref{eq:form3} are trivially equivalent as it  can be readily noted by subtracting  the two series above:\\
\begin{footnotesize}
\begin{equation}\label{form4}
 SUM1-SUM2 =  \sum_1^{\infty} \frac{(2k)^2 -1 }{(2k-1)^2 (2k+1)^2 } = \sum_1^{\infty} \frac{1 }{(2k-1) (2k+1) } =  \frac{1}{2}[  \sum_1^{\infty} \frac{1 }{(2k-1) }  -\sum_1^{\infty} \frac{1}{(2k+1) }] =\frac{1}{2}\\
\end{equation}
\end{footnotesize}

\noindent
Further, it follows from formula \ref{eq:form1},\\

\begin{footnotesize}
\begin{equation}\label{form5}
\frac{\pi^2}{4} = 1+ \sum_{k=1}^{k=\infty}  \frac{1}{(2k+1)^2}+\sum_{k=1}^{k=\infty}  \frac{1}{(2k-1)^2} =1 + \sum_{k=1}^{k=\infty}\frac{(8k^2+2)}{ (2k-1)^2 (2k+1)^2} = 1+ 2 \cdot SUM1 +2 \cdot SUM2 = 4 \cdot SUM1  \\
\end{equation}
\end{footnotesize}

\noindent
Thus indeed the formula \ref{eq:form1} is equivalent to $4 \cdot SUM1=\frac{\pi^2}{4}$ and, in  consequence, to 
the  formulas \ref{eq:form2} and \ref{eq:form3}:\\

\noindent
\section{Appendix A2}
\noindent
\noindent
The functions of the form:\\

\begin{equation}\label{form7}
  \Phi^+_n(x) = \frac{i}{\sqrt{L}} exp({\frac{+in\pi x}{L}}), \Phi^-_n (x) = \frac{i}{\sqrt{L}} exp({\frac{-in\pi x}{L}})\\
\end{equation}

\noindent
are eigenfunctions of the momentum operator, but they do not 
form an orthonormal set. The usual procedure is  to propose  an orthonormal
set of momentum eigenfunctions  normalized in the well,
$ \Xi(x) = \frac{1}{\sqrt{L}} exp({\frac{ipx}{\hbar}})$,  
which fulfill periodic boundary
conditions, $ \Xi(L)= \Xi(0)$.  \\

\noindent
These momentum operator eigenfunctions  are indeed  orthonormal  in the range $[0,L]$ as basic explicit calculation shows:\\

\noindent
$\int_0^L (\Xi^+_n(x))^* \Xi^+_m(x) dx = \frac{1}{{L}} \int_0^L exp({\frac{-i2n\pi x}{L}})exp({\frac{i2m\pi x}{L}})dx$\\

\noindent
for $ m \neg n$ one has:\\

\noindent
$\frac{1}{{L}} \int_0^L exp({\frac{-i2n\pi x}{L}})exp({\frac{i2m\pi x}{L}})dx = \frac{1}{i2(m-n)\pi}[exp({i2(m-n)\pi})-1]=0$\\

\noindent
for $ m = n$ one  gets $1$. Functions with minus sign  are complex-conjugates of the functions with the sign, $(+)$,
thus the orthonormality is also
valid for them. The scalar product of  functions with different signs should always give zero:\\

\noindent
$\int_0^L (\Xi^-_n(x))^* \Xi^+_m(x) dx = \frac{1}{{L}} \int_0^L exp({\frac{i2n\pi x}{L}})exp({\frac{i2m\pi x}{L}})dx$\\  

\noindent
$\frac{1}{{L}} \int_0^L exp({\frac{i2n\pi x}{L}})exp({\frac{i2m\pi x}{L}})dx = \frac{1}{i2(m+n)\pi}[exp({i2(m+n)\pi})-1]=0$\\

\end{document}